\documentclass[a4paper]{jpconf}
\usepackage{graphicx}
\begin{document}

\title{Internal convection in thermoelectric generator models}

\author{Y Apertet$^1$, H Ouerdane$^2$, C Goupil$^{3,4}$ and Ph Lec{\oe}ur$^1$}
\address{$^1$ Institut d'Electronique Fondamentale, Universit\'e Paris-Sud, CNRS, UMR 8622, F-91405 Orsay, France}
\address{$^2$ CNRT Mat\'eriaux UMS CNRS 3318, 6 Boulevard Mar\'echal Juin, 14050 Caen Cedex, France}
\address{$^3$ Laboratoire CRISMAT, UMR 6508 CNRS, ENSICAEN et Universit\'e de Caen Basse Normandie, 6 Boulevard Mar\'echal Juin, F-14050 Caen, France}
\address{$^4$ To whom correspondence should be sent.}

\ead{christophe.goupil@ensicaen.fr}

\begin{abstract}
Coupling between heat and electrical currents is at the heart of thermoelectric processes. In a thermoelectric system this may be seen, from a thermal viewpoint, as an additional thermal flux linked to the appearance of an electrical current. Since this additional flux is associated with the global displacement of charge carriers in the system, it can be qualified as convective in opposition to the conductive part related to both phonon transport and heat transport by electrons under open circuit condition as, e.g., in the Wiedemann-Franz relation. In this article we demonstrate that considering the convective part of the thermal flux allows both new insight into the thermoelectric energy conversion and the derivation of the maximum power condition for generators with realistic thermal coupling.
\end{abstract}

\section{Introduction}
The field of thermoelectricity was born in the 1820s after Seebeck reported his observation of a phenomenon he interpreted as thermomagnetism. Orsted, in the same years, gave a better interpretation of the phenomenon and termed it thermoelectricity: the appearance of an electrical current in a circuit subjected to a temperature difference. The name of Seebeck was later given to the coefficient linking the temperature difference $\Delta T$ and the resulting electromotive force $\Delta V$: $\alpha = - \Delta V / \Delta T$. A few years after Seebeck's works were published~\cite{Seebeck1826}, Peltier noticed that, conversely, an electrical current $I$ may absorb or release heat at a junction between two different materials \cite{Peltier1834}. The flux of absorbed heat $I_Q$ is related to the electrical current $I$ through the difference of Peltier coefficients $\Pi$ between the materials of the junction: $I_Q = (\Pi_b - \Pi_a) I$ (with $I$ positive when flowing from material $a$ to material $b$). The relation between these two effects was later clearly expressed by Thomson who established the connection between the Peltier and Seebeck coefficients: $\Pi = \alpha T$, with $T$ the temperature of the considered  material \cite{Thomson1851}. This relation demonstrates that both effects have the same physical origin: the ability of charge carriers to carry heat along with electrical charge. As thermoelectric (TE) devices possess the particularity to strongly couple heat and electrical transports, they can be used in refrigeration or energy conversion applications \cite{Rowe2006}. However to achieve good performances both TE materials and device design have to be optimized. For materials, optimization amounts to increasing the so-called figure of merit $ZT$ (see for example the recent review by Shakouri \cite{Shakouri2011}). Yet, to make the best use of the materials it is necessary also to lead a reflection on the integration of the devices into an external environment. One thus needs a complete model of thermoelectric device to fully comprehend the interaction between the constitutive laws of the device and the laws related to the external exchanges.

In this work we propose a simple model of thermoelectric generator (TEG) that accounts for the heat transported by electrical carriers; this results in the introduction of an effective thermal conductance associated with this phenomenon. As this heat propagation is associated with global movement of particles, by analogy with fluid mechanics, we call it \emph{convection}. We then study and illustrate the consequences that this additional term may bear on both the physics of TE energy conversion and on the practical design and optimization of thermoelectric generators (TEGs). 

\section{Thermoelectric modeling: Accounting for a convective term}

\begin{figure}
	\centering
		\includegraphics[width=1\textwidth]{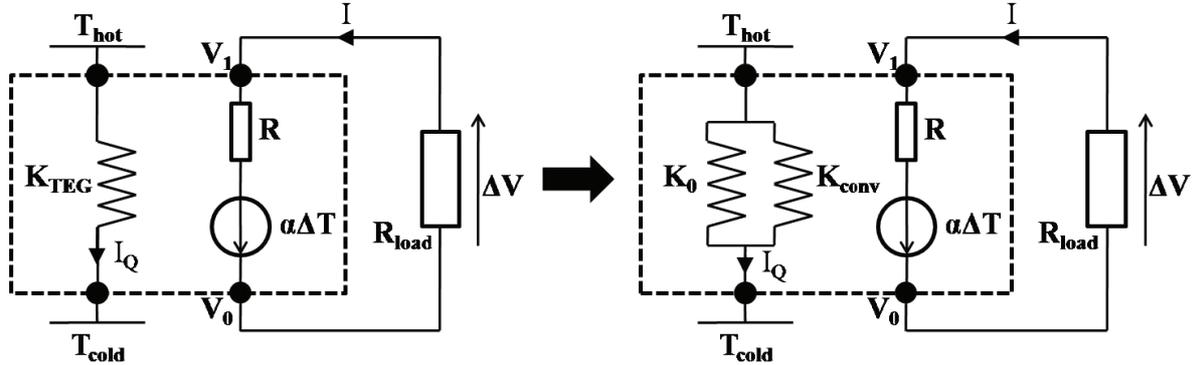}
	\caption{Thermoelectric generator: Decomposition of the TEG thermal conductance into a conductive part $K_0$ and a convective part $K_{\rm conv}$.}
	\label{fig:figure1}
\end{figure}

Let us consider a thermoelectric generator connected to two heat baths at temperature $T_{\rm hot}$ and $T_{\rm cold}$ ($T_{\rm hot} > T_{\rm cold}$) respectively with an average temperature $T = (T_{\rm hot} +  T_{\rm cold}) / 2$. Though a thermoelectric device may operate indifferently either as a cooler or as a generator, in this work, we restrict our analysis to the case of a generator to which an electrical load resistance $R_{load}$ is connected in order to extract power. Now that we have set the thermal and electrical environment, we state the phenomenological laws that govern the device by relating the electrical current $I$ and the average thermal current $I_Q$ to the voltage $\Delta V$ and the temperature difference $\Delta T$ across the generator. The equations follow directly from the force-flux formalism introduced by Onsager and then extended to thermoelectricity by Callen \cite{Callen1948}. For our purpose, we find convenient to use the macroscopic coefficients and potential differences, rather than local coefficients and gradients, and we get:

\begin{equation} \label{eq:I}
I = \frac{1}{R}\Delta V + \frac{\alpha}{R} \Delta T
\end{equation}
\begin{equation} \label{eq:IQ}
I_Q = \frac{\alpha T}{R} \Delta V + \left(\frac{\alpha^2 T}{R} + K_{0}\right) \Delta T
\end{equation}

\noindent where $R$ is the electrical resistance, $K_0$ the thermal conductance under open-circuit condition and $\alpha$ the Seebeck coefficient. The TEG is also characterized by its figure of merit $ZT$ obtained from these three parameters: $ZT = \alpha^2 T /R K_0$. This dimensionless number gives a direct quantitative estimation of the generator's quality \cite{Ioffe1957}. The convention for the currents and potential differences, $\Delta V = V_1 - V_0$ and $\Delta T = T_{\rm hot} - T_{\rm cold}$, are displayed in Fig.~(\ref{fig:figure1}).

Equation~(\ref{eq:I}) gives a definition of the Seebeck coefficient under open-circuit condition ($I = 0$), namely $\alpha = - \Delta V / \Delta T$. If the Seebeck coefficient $\alpha$ vanishes, thermoelectric effects become negligible, and one recovers the traditional macroscopic expressions of Ohm's law and Fourier's law, from Eqs.~(\ref{eq:I}) and (\ref{eq:IQ}) respectively.

The two previous equations may be combined to express the thermal flux as the sum of two terms:
\begin{equation}\label{IQ1}
I_Q = \alpha T I + K_{0} \Delta T
\end{equation}

\noindent The first term is the contribution due to the heat transported within the electrical current, the so-called \emph{convective} thermal flux \cite{Pottier2007,Saleh1991}, and the second is the conductive part traditionally associated with Fourier's law. The quantity of energy transported by each carrier is $|\alpha| T e$, $e$ being the elementary electric charge.

To compare these two terms we now rewrite Eq.~(\ref{IQ1}) to obtain a form similar to Fourier's law: A relation of proportionality between the heat current and the temperature difference. For this purpose we express the electrical current through the resistive load as a function of $\Delta T$. From Ohm's law: $I = -\Delta V / R_{\rm load}$, we get:

\begin{equation}\label{Int1}
I = \frac{\Delta V + \alpha \Delta T}{R} = \frac{\alpha \Delta T}{R_{\rm load} + R}
\end{equation}

\noindent After substitution of the above expression for $I$ in Eq.~(\ref{IQ1}), the average heat flux may be expressed as:

\begin{equation}\label{IQ2}
I_Q = \left( \underbrace{\frac{\alpha^2 T}{R_{\rm load} + R}}_{\displaystyle K_{\rm conv}} + K_{0} \right)\Delta T = K_{\rm TEG}\Delta T,
\end{equation}

\noindent where an additional term, $K_{\rm conv}$, for the total thermal conductance of the generator now appears clearly. This term is directly related to the electrical condition on the circuit since the load resistance $R_{\rm load}$ is explicitly involved in its definition. The effective thermal conductance of the generator $K_{\rm TEG}$ is thereby equivalent to the two thermal conductances $K_{\rm conv}$ and $K_{0}$ in parallel as illustrated in Fig.~(\ref{fig:figure1}).
When $R_{\rm load} \rightarrow \infty$, i.e., in open-circuit condition, we recover, as expected, the fact that $K_{\rm TEG}$ is equivalent to $K_{0}$ only, whereas when $R_{\rm load} \rightarrow 0$, $K_{\rm TEG} = \alpha^2 T / R + K_{0}$.

To derive Eq.~(\ref{IQ2}) we made a few assumptions: We considered only an average value for the heat flux; we neglected the Joule effect and the thermal energy converted into electric energy along the device. These approximations are reasonable if the temperature difference $\Delta T$ is kept small; a condition that must be satisfied to make use of constant coefficients to describe the TE generator. All these assumptions amount to stating that we only consider a linear framework. Second, we neglected the heat flowing through the electrical load. So, dealing with a true unileg generator as in Ref.~\cite{Lemonnier2008}, implies $K_{\rm TEG} \gg K_{\rm load}$. This configuration may also be related to a generator with two legs: The parameters then used are only equivalent ones \cite{Ioffe1957}, in which case the load sustains no temperature difference, so that no heat flux goes through it.

\section{Illustrative examples of the effects of thermoelectric convection}
In this section we illustrate the impact that the convective part of the generator thermal conductance may have on the properties and operation of a TEG. We first look at the construction and meaning of the figure of merit $ZT$; next we discuss the question of optimization of thermoelectric generators when the connections to the heat reservoirs are not perfect; finally we examine the effective thermal conductance of two thermoelectric modules in parallel.  

\subsection{On the meaning of the figure of merit ZT}
To discuss the properties of thermoelectric systems, we find convenient to proceed by analogy using the classical heat transfer description in terms of convection and conduction. To determine whether heat transfer in a fluid is dominated either by conduction or by convection, one may introduce the Prandtl number $\sigma_{\rm p}$ defined as

\begin{equation}\label{Prandtl}
\sigma_{\rm p} = \frac{\nu}{D} 
\end{equation}

\noindent where $\nu$ is the kinematic viscosity of this fluid, characterizing momentum diffusion, and $D$ its thermal diffusivity, characterizing heat diffusion \cite{Blundell2010}. We retained this idea to define a \emph{thermoelectric Prandtl number} $(\sigma_{\rm p})_{_{\rm TE}}$ given by the ratio of the convective thermal flux and of the conductive thermal flux:

\begin{equation}\label{PrandtlTE}
(\sigma_{\rm p})_{_{\rm TE}} = \frac{K_{\rm conv} \Delta T}{K_{0} \Delta T} = \frac{\alpha^2 T}{K_{0}(R + R_{\rm load})} = \frac{ZT}{1 + R_{\rm load} / R}
\end{equation}

\noindent It appears that $(\sigma_{\rm p})_{_{\rm TE}}$ is proportional $ZT$. The proportionality coefficient depends on the ratio between the electrical resistances of the generator and the load. If the electrical circuit is shorted, i.e., if $R_{\rm load} \ll R$, the thermoelectric Prandtl number $(\sigma_{\rm p})_{_{\rm TE}}$ is then equal to $ZT$, which is the maximum value of the ratio between the convective part and the conductive part of the thermal flux. It is consistent with the fact that $ZT$ must be increased to improve the TEG performances: The convective thermal flux is a direct consequence of the coupling between heat and electrical charge transport, and, as such, it is a necessary process to obtain energy conversion. Conversely, the conductive thermal flux amounts to pure loss effects only: So maximizing $ZT$ is equivalent to maximizing the \emph{useful part} of the thermal flux with respect to the losses.

The ratio $R_{\rm load} / R$ controls the convective component of the heat flux. However, maximization of $(\sigma_{\rm p})_{_{\rm TE}}$ does not lead to the best performances: To extract power from the heat that flows in the generator, one has to satisfy impedance matching too. If $R_{\rm load}$ becomes too small, the electrical current, and hence the convective thermal current, will increase but the power will decrease. It appears that the optimal value of $(\sigma_{\rm p})_{_{\rm TE}}$ associated to the maximum power output is not $ZT$ but $ZT /2$. We can even extend this analysis to the maximization of the TEG efficiency: As shown by Ioffe \cite{Ioffe1957}, for a TEG with constant parameters the efficiency is maximized when $R_{\rm load} / R = \sqrt{1 + ZT}$. Thereby, for the efficiency the optimal value of $(\sigma_{\rm p})_{_{\rm TE}}$ is $ ZT / (1 + \sqrt{1 + ZT})$.

This last result should be compared with the compatibility approach of thermoelectric optimization: Snyder and Ursell \cite{Snyder2003} demonstrated that in order to obtain the best efficiency, the relative current $u = J / (\kappa \nabla T)$ (with $J$ the electrical current density and $\kappa$ the local thermal conductivity under open circuit-condition) has to be kept constant along the device and should be equal to an optimal value $s = (\sqrt{1 + ZT} - 1) / \alpha T$. Expressed differently the condition for efficiency maximization reads: $\alpha T J / (\kappa \nabla T) = (\sqrt{1 + ZT} - 1)$. As the term on the left hand side is a local definition of $(\sigma_{\rm p})_{_{\rm TE}}$, we recover exactly the same result as above. It is interesting to note that similar conditions were derived using the ratio of the total heat current to the convective thermal flux in Refs. \cite{Clingman1961, Goupil2011}.

Finally the notion of convective thermal conductance can also be useful to practically determine the figure of merit of a thermoelectric generator: In 1966, Lisker proposed to obtain $ZT$ by measuring the effective thermal conductance of the TE module for two different electrical load conditions instead of extracting from three different measurements the value of each parameter $R$, $\alpha$ and $K_0$ \cite{Lisker1966}. The ratio $K_{\Delta V = 0}/K_0$ is indeed equal to $ZT + 1$.

\subsection{Thermal matching of thermoelectric generators}
\begin{figure}
	\centering
		\includegraphics[width=1\textwidth]{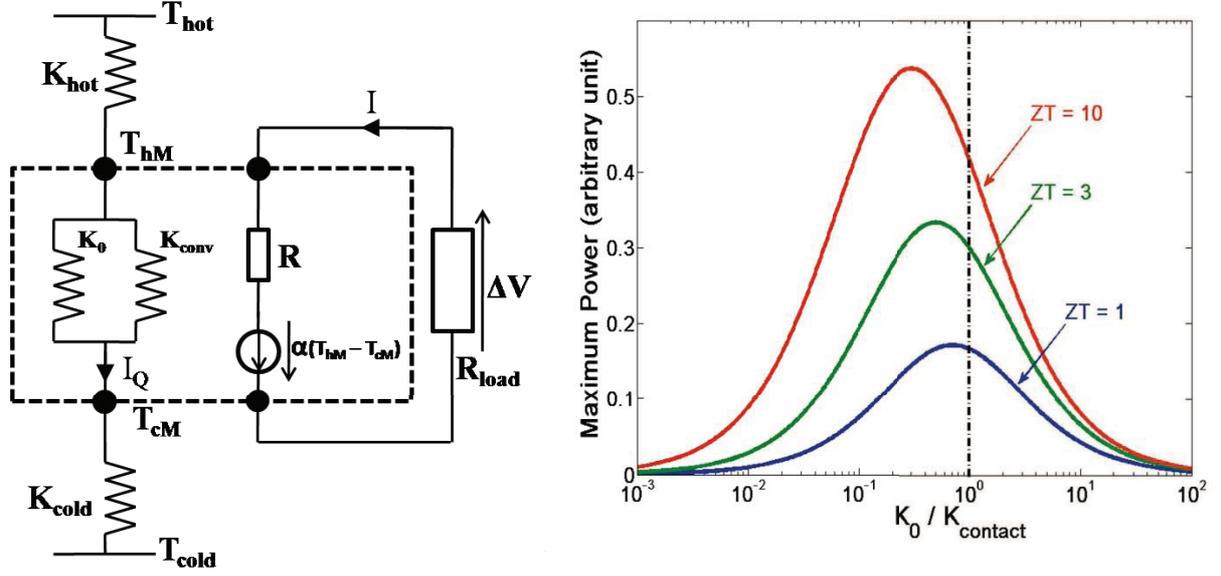}
	\caption{Thermoelectric generator with dissipative thermal coupling (left) and maximum power output as a function of the ratio $K_0 / K_{\rm contact}$ for different values of ZT (right).}
	\label{fig:figure2}
\end{figure}

Assume that the generator is not directly coupled to the heat reservoirs; the finite thermal conductance of the contacts must be taken into account. These are denoted $K_{\rm hot}$ on the hot side and $K_{\rm cold}$ on the cold side of the generator. The total equivalent conductance for both contacts is given by $K_{\rm contact}^{-1} = K_{\rm hot}^{-1} + K_{\rm cold}^{-1}$. The main effect of these imperfect contacts is that the actual temperature difference across the TEG is not $\Delta T$ but $\Delta T' = T_{\rm hM} - T_{\rm cM}$ as illustrated on the left side of Fig.~(\ref{fig:figure2}).

We first need to determine the relation between these two temperature differences. To do so we use an analogue to the potential divider formula for the temperature, assuming that the heat flux remains constant along the structure (this hypothesis is the same as that used to derive the expression of $K_{\rm conv}$). We obtain the following relation:

\begin{equation}\label{kapctc}
\Delta T' = T_{\rm hM} - T_{\rm cM} \approx \frac{K_{\rm contact}}{K_{\rm TEG}+K_{\rm contact}}~\Delta T
\end{equation}

\noindent Since the thermal conductance $K_{\rm TEG}$ depends on the electrical load resistance through $K_{\rm conv}$, the temperature difference across the generator, $\Delta T'$, is modulated by the electrical load condition. The consequence of this modulation is that the electrical circuit modeling of the TEG as diplayed in Fig.~(\ref{fig:figure2}) is no longer a Th\'evenin generator: In such model the voltage source must be independent of the load which is not the case when finite thermal contacts are introduced. To keep using a proper Th\'evenin generator we split the thermoelectric voltage $V_{\rm oc} = \alpha \Delta T'$ into two terms: One is independent of the electrical load, and the other is linked to an internal resistance~\cite{Apertet2012a}:

\begin{equation}\label{voc}
V_{\rm oc} =\alpha \Delta T \frac{K_{\rm contact}}{K_{0}+K_{\rm contact}}-I \frac{\alpha^2 T}{K_{_{I=0}}+K_{\rm contact}}
\end{equation} 

\noindent This equation is of the form: $V_{\rm oc} = V_{\rm oc}' - IR'$ with $R'= \alpha^2 T / (K_{_{I=0}}+K_{\rm contact})$. We thus obtain a rigorous Th\'evenin modeling of the electrical part of the TEG with the definitions of the open circuit voltage given by $V_{\rm oc}'$ and the internal resistance is $R_{\rm TEG} = R + R'$: As for the effective thermal conductance, the effective electrical resistance of the generator shows an additional contribution directly related to the thermal conditions imposed on the generator through the thermoelectric coupling coefficient $\alpha$. Further, it is remarkable that the expressions of $R'$ and $K_{\rm conv}$ present the same form. One can switch from one to the other by changing thermal conductances to electrical resistances and \emph{vice versa}. Recently, the introduction of an extra electrical resistance depending on thermal conditions was also suggested by Spry \cite{Spry2012}. With our model, the power produced by the TEG may thus be simply expressed as:

\begin{equation}\label{Pprod1}
P = \Delta VI = \frac{{V_{\rm oc}'}^2 R_{\rm load}}{(R_{\rm TEG}+R_{\rm load})^2}
\end{equation}

\noindent from which we deduce that power maximization is obtained for $R_{\rm load} = R_{\rm TEG}$. Such a condition was already derived mathematically by Freunek and co-workers \cite{Freunek2009}.

As regards thermal matching, Stevens proposed that to maximize power production, the TEG should be thermally matched with $K_{0} = K_{\rm contact}$. Let us examine the situation using the equations we just derived. An expression for the maximum power may be obtained using Eq.~(\ref{Pprod1}) in the case of electrical matching and the definition of $R_{\rm TEG}$: 

\begin{equation}\label{pmax}
P_{\rm max} = \frac{(K_{\rm contact}\Delta T)^2}{4(K_{_{I=0}}+K_{\rm contact})T}~\frac{ZT}{1+ZT+K_{\rm contact}/K_{_{I=0}}}
\end{equation}

\noindent For a given value of $ZT$, $P_{\rm max}$ shows a clear dependence on the ratio $K_{_{I=0}} / K_{\rm contact}$. If we consider a fixed value of $K_{\rm contact}$, due for example to technological constraints, the choice of the value for $K_{_{I=0}}$ will strongly impact the ability of the generator to produce power. On the right side of Fig.~(\ref{fig:figure2}) we show the relation between $P_{\rm max}$  and $K_{_{I=0}} / K_{\rm contact}$ for three different values of $ZT$ (1, 3 and 10). Note that for each point electrical impedance matching is achieved. We see that this condition for power maximization through thermal impedance matching slightly differs from the one obtained by Stevens, and that it depends on $ZT$. Actually the optimum value of $K_0$ corresponds to the thermal matching between $K_{\rm contact}$ and $K_{\rm TEG}$: $K_{\rm TEG} = K_{\rm contact}$ \cite{Apertet2012a}. The dependence on $ZT$ is explained by the fact that the greater $ZT$ is, the greater the contribution of $K_{\rm conv}$ to $K_{\rm TEG}$ is. Thus for sufficiently high TEG efficiency, i.e., high $ZT$, the discrepancy between these two conditions of thermal impedance matching is not negligible: See the difference in Fig.~(\ref{fig:figure2}) between the maximum power and the value of power for the condition $K_{_{I=0}} = K_{\rm contact}$ displayed by the vertical dot-dashed line.

Finally we express the simultaneous conditions of thermal and electrical matching in terms of ratios $K_{0} / K_{\rm contact}$ and $R_{\rm load} / R$ (as $R$ and $K_{0}$ are more suitable quantities for device design than $K_{\rm TEG}$ and $R_{\rm TEG}$):

\begin{equation}
\frac{K_{0}}{K_{\rm contact}} = \frac{1}{\sqrt{1 + ZT}}
\end{equation}
\begin{equation}
\frac{R_{\rm load}}{R} = \sqrt{1 + ZT}
\end{equation}

Once again, we emphasize the symmetry that arises between thermal and electrical quantities. In a recent publication, Yazawa and Shakouri obtained the same relations from a more mathematical viewpoint~\cite{Yazawa2012}.

\subsection{Effective thermal conduction increase in inhomogeneous material \cite{Apertet2012b}}

\begin{figure}
	\centering
		\includegraphics[width=0.7\textwidth]{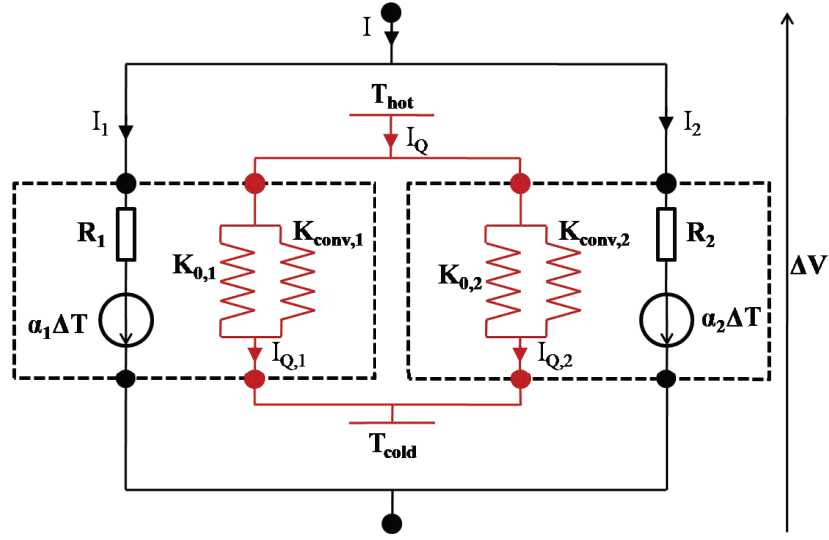}
	\caption{Association of 2 thermoelectric generators in parallel.}
	\label{fig:figure3}
\end{figure}

Our last example concerns the determination of the effective thermal conductance of two thermoelectric modules in parallel as represented in Fig.~(\ref{fig:figure3}). We keep the same notations for modules' properties except that we add a subscript ($1$ or $2$) to distinguish the two modules. First, we derive the total average thermal flux $I_{Q_{_{\rm eq}}}$ inside the system: It is given by the sum of the average thermal flux for each module, with a conductive and a convective part for each. So we have:

\begin{equation}
\label{eq:conservationIQ}
I_{Q_{_{\rm eq}}}=I_{Q_{_{1}}}+I_{Q_{_{2}}} = \alpha_1 T I_1 + K_{0,1}\Delta T + \alpha_2 T I_2 + K_{0,2}\Delta T
\end{equation}

As we focus on the open electrical circuit condition to determine the equivalent thermal conductance $K_{0,{\rm eq}}$, the total electrical current is set to $I = 0$. However this does not mean that the currents $I_1$ and $I_2$ vanish as well: The only consequence is that $I_1 = - I_2$. A simple analysis of the electrical circuit in Fig.~(\ref{fig:figure3}) leads to the following expression for the internal current:
\begin{equation}
\label{eq:I1}
I_1 = - I_2 = \frac{(\alpha_1 - \alpha_2)\Delta T}{R_1 + R_2}
\end{equation}

This internal current only exists if the electromotive forces, given by $\alpha \Delta T$, do not compensate each other, i.e., if $\alpha_1 \neq \alpha_2$. In that case we have to consider the convective contribution to the thermal flux. Substituting Eq.~(\ref{eq:I1}) in Eq.~(\ref{eq:conservationIQ}) and simplifying by $\Delta T$ we get the equivalent open circuit thermal conductance:
\begin{equation}
\label{eq:condthermeff}
K_{0,{\rm eq}} = K_{0,1} + K_{0,2} + \frac{(\alpha_1 - \alpha_2)^2T}{R_1 + R_2}
\end{equation}

This last equation demonstrates that the convective component of the thermal flux has a significant impact on the effective thermal conductance of the two thermoelectric modules in parallel if they possess very dissimilar Seebeck coefficients. However this term is of limited interest if the electrical resistances of the modules are too high: the magnitude of electrical internal current is then small and, consequently, so is the convective thermal flux. Knowledge of the effective thermal conductance may be useful when dealing with superlattices for which the layers can be viewed as parallel modules \cite{Saleh1991}. More surprisingly, Price obtained a similar expression for the effective thermal conductance of an ambipolar conductor where the subscripts 1 and 2 are associated with electrons and holes repectively: The two types of carriers are indeed considered as evolving in parallel inside the system \cite{Price1955}.

\section{Conclusion}

Convective thermal flux will have more and more impact on the behavior of thermoelectric generator as the performance of materials will be increased. Indeed we have shown that the proportion of convective to conductive thermal current is directly related to the figure of merit $ZT$ which is expected to increase significantly in the future to allow a wider use of thermoelectric power production. In this paper we have demonstrated that this additional contribution to heat flow should not be neglected for device design as well as for materials research, especially those concerning composites where internal currents can develop and lead to a performance decrease.

\section*{Acknowledgments}
This work is part of the SYSPACTE projects funded by the Fonds Unifi\'e Interminist\'eriel 7. Y. A. acknowledges financial support from the Minist\`ere de l'Enseignement Sup\'erieur et de la Recherche.

\end{document}